**An Experimental and Theoretical Investigation of Spray Characteristics of Impinging Jets in Impact Wave Regime.**

N.S. Rodrigues, V. Kulkarni, J. Gao, J. Chen, P.E. Sojka

School of Mechanical Engineering, Purdue University

West Lafayette, IN 47906

Abstract:

The current study focuses on experimentally and theoretically improving characterization of the drop size and drop velocity for like-on-like doublet impinging jets. The experimental measurements were made using Phase Doppler Anemometry (PDA) at jet Weber numbers $We_j$ corresponding to the impact wave regime of impinging jet atomization. A more suitable dynamic range was used for PDA measurements compared to literature, resulting in more accurate experimental measurements for drop diameters and velocities. There is some disagreement in literature regarding the ability of linear stability analysis to accurately predict drop diameters in the impact wave regime. This work seeks to provide some clarity. It was discovered that the assumed uniform jet velocity profile was a contributing factor for deviation between diameter predictions based on models in literature and experimental measurements. Analytical expressions that depend on parameters based on the assumed jet velocity profile are presented in this work. Predictions based on parabolic and 1/7th power law turbulent profiles were considered and show better agreement to the experimental measurements compared to predictions based on the previous models. Experimental mean drop velocity measurements were compared with predictions from a force balance analysis and it was observed that the assumed jet velocity profile also influences the predicted velocities, with the turbulent profile agreeing best with experimental mean velocity. It is concluded that the assumed jet velocity profile has a predominant effect on drop diameter and velocity predictions.

Corresponding Authors: Neil S. Rodrigues (neilrodrigues@asme.org), Varun Kulkarni (varun14kul@gmail.com).

## LIST OF SYMBOLS

| | |
|---|---|
| $b^*$ | Dimensionless distance from center of jet to separation point [-] |
| $d_D$ | Predicted drop diameter [µm] |
| $d_j$ | Jet diameter [mm] |
| $d_0$ | Orifice diameter [mm] |
| $D_{10}$ | Arithmetic mean drop diameter [µm] |
| $D_{32}$ | Sauter mean drop diameter [µm] |
| $f_0$ | Number pdf [µm$^{-1}$] |
| $f_2$ | Area pdf [µm$^{-1}$] |
| $f_3$ | Volume pdf [µm$^{-1}$] |
| $Fr_j$ | Jet Froude number [-] |
| $K^*$ | Dimensionless sheet thickness parameter [-] |
| MMD | Mass Median Diameter [µm] |
| $q^*$ | Dimensionless radial distance from the separation point [-] |
| $q_j^*$ | Dimensionless location of the jet interface [-] |
| $R_j$ | Jet radius [mm] |
| $Re_D$ | Drop Reynolds number [-] |
| $Re_j$ | Jet Reynolds number [-] |
| s | Ratio of ambient gas density to liquid density [-] |
| $U_j$ | Jet velocity [m·s$^{-1}$] |
| $U_d$ | Drop velocity [m·s$^{-1}$] |
| $U_{z-mean}$ | Experimentally measured mean drop velocity [m·s$^{-1}$] |
| $We_d$ | Drop Weber number [-] |
| $We_j$ | Jet Weber number [-] |
| $L/d_0$ | Internal length to orifice diameter ratio [-] |
| $x/d_0$ | Free jet length to orifice diameter ratio [-] |
| α | Ratio of sheet velocity to jet velocity [-] |
| γ | Liquid/Gas surface tension [N/m] |
| θ | Half-impingement angle [°] |
| $\mu_l$ | **Liquid viscosity [Pa-s]** |
| $\rho_g$ | **Ambient gas density [kg-m$^{-3}$]** |
| $\rho_l$ | **Liquid density [kg-m$^{-3}$]** |
| $\phi$ | **Sheet azimuthal angle [°]** |

**1 Introduction.**

Common injectors for liquid rocket engines include the impinging jet injector (Anderson et al. 1995), coaxial injector (Vingert et al. 1995), and pintle injector (Heister 2004). Among these, the impinging jet injector is considered to be the most popular due to ease in fabrication, desirable atomization characteristics, and high performance mixing. An impinging jet atomizer with two jets of the same liquid (either fuel or oxidizer) is called a like-on-like doublet. In contrast, an unlike doublet consists of one jet of fuel and one jet of oxidizer (Gill and Nurick 1976).

Since the injection, atomization, mixing, and combustion occur nearly simultaneously in a fuel/oxidizer system, uncoupling is required to separately investigate each of these processes (Humble et al. 1995). For instance, in liquid impinging jet atomization studies, water has been traditionally used instead of the propellant to eliminate the combustion process.

The dimensionless parameter that is typically used to characterize impinging jet atomization is the jet Weber number $We_j$,

$$We_j = \frac{\rho_l U_j^2 d_j}{\gamma}, \tag{1}$$

which represents the ratio of the inertial force to the surface tension force. In the above equation $\rho_l$ is the liquid density, $U_j$ is the jet velocity, $d_j$ is the jet diameter (assumed to be equal to the orifice diameter $d_0$), and $\gamma$ is the liquid-gas surface tension. Another parameter that influences the atomization process is the ratio of ambient gas density to liquid density $s$,

$$s = \frac{\rho_g}{\rho_l}, \tag{2}$$

where $\rho_g$ denotes the ambient gas density. The flow behavior of the jets (laminar or turbulent) is also of importance and is characterized by the jet Reynolds number $Re_j$,

$$Re_j = \frac{\rho_l U_j d_j}{\mu_l}, \tag{3}$$

which represents the ratio of the inertial force to the viscous force. In the above equation $\mu_l$ is the liquid viscosity.

Representative diameters typically used to characterize the impinging jet spray include the arithmetic mean diameter $D_{10}$, Sauter mean diameter $D_{32}$, and the mass median diameter $MMD$. The arithmetic mean diameter is a first order mean and is used widely for comparison. The Sauter mean diameter is a fifth order mean that represents the volume to surface area ratio. The mass median diameter is defined as a representative diameter such that 50% of total liquid volume is in drops of smaller diameters. The expressions to calculate these diameters are:

$$D_{10} = \frac{\sum N_i D_i}{\sum N_i}, \tag{4}$$

$$D_{32} = \frac{\sum N_i D_i^3}{\sum N_i D_i^2}, \tag{5}$$

$$0.5 = \int_0^{MMD} f_3(D) \, dD. \tag{6}$$

$N_i$ and $D_i$ are the number of drops and the drop diameters respectively for the mean diameter calculations. $D_{32}$ is the most applicable mean diameter for combustion purposes because it best describes the fineness of the spray. In equation 6, $f_3$ symbolizes the volume probability density function (pdf). The volume pdf is the probability density that a drop has a volume of $\pi D^3/6$. Two other methods to quantify the polydisperse nature of the impinging jet spray is the number pdf $f_0$ (probability density that a drop has diameter $D$), and the area pdf $f_2$ (probability density that a drop has surface area $\pi D^2$) (Lefebvre 1989).

The atomization mechanism for the impinging jet injector with a like-on-like doublet configuration involves two cylindrical liquid jets with equal jet velocity and jet diameter impinging each other. This impingement first creates a flat liquid sheet that is perpendicular to the momentum vectors of the two jets. Instabilities are present on this liquid sheet that promote breakup. The general breakup behavior is the sheet disintegrating into ligaments, and the ligaments then disintegrating into drops. Figure 1 provides a schematic of this breakup process. Disintegration is primarily caused by the aerodynamic and inertial forces and is opposed by the surface tension force. The liquid viscosity acts as a dampener for breakup – liquids possessing higher viscosity tend to break up into relatively larger drops. Since water is typically used as the test liquid for impinging jet studies, an inviscid analysis sufficiently describes the physics.

Breakup patterns in the three regimes of impinging jet atomization evolve as the jet Weber number and the jet Reynolds number increase. However, clear boundaries for breakup patterns are not reported due to the influence of injector geometry, among other factors. The following approximate conditions are summarized from Anderson et al. (2006) and von Kampen et al. (2006). The laminar sheet regime is observed at lower jet Weber and jet Reynolds numbers. Patterns in this regime include the closed rim with droplet formation pattern ($We_j$ = 160, $Re_j$ = 2,480), the open rim pattern ($We_j$ = 320, $Re_j$ = 3,450), and the rimless separation pattern ($We_j$ = 670, $Re_j$ = 5,020). The impact wave regime occurs at higher jet Weber and Reynolds numbers. Patterns in this regime include the ligament structures pattern ($We_j$ = 3,420, $Re_j$ = 10,030) and the fully developed pattern ($We_j$ = 19,650, $Re_j$ = 23,740). Notable features of this regime include impact waves emitting from the impingement point and a large number of small drops. The aerodynamic breakup and atomization regime is also observed but only at conditions of high ambient gas density. Only the impact wave and the aerodynamic breakup and atomization regimes occur in practical liquid rocket engines.

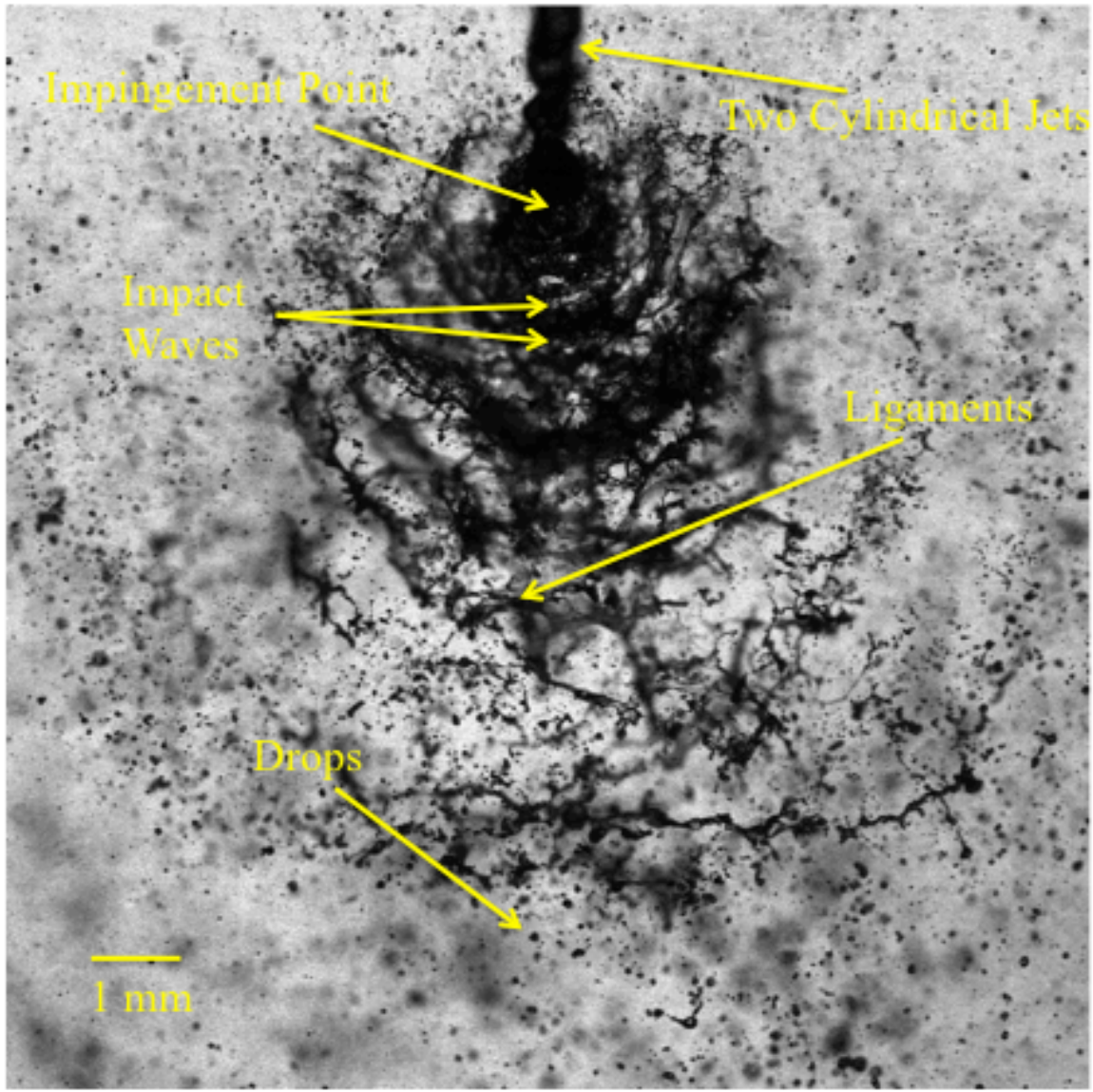

**Figure 1.** Impinging jet spray formation.

Previous experimental studies have investigated scaling relationships between the inertial force (often controlled using the jet velocity) and spray characteristics such as frequency of waves, sheet breakup length, drop diameter, and drop velocity. The study by Heidmann et al. (1957) was one of the earliest works that studied the frequency of waves; frequency was observed to increase with jet velocity. Anderson et al. (1992) studied the sheet breakup length; sheet breakup length is observed to decrease with an increase in the jet velocity.

Dombrowski and Hooper (1963) studied the effect of varying jet velocity on the $D_{32}$ mean diameter for both laminar and turbulent jets. In the turbulent case, the mean drop size was observed to decrease with increasing jet velocity. The drop size was quantified in the central region of the spray using still photography. Still photography using the shadowgraphy technique with image analysis has also been extensively applied in other impinging jet atomization studies. However, as a two-dimensional measurement technique, it only resolves characteristics in/near the focal plane of the imaging lens. Furthermore, small drops are typically not measured due to the limited dynamic range. Images are usually analyzed in the plane of the sheet formed by the two jets.

The most widely referenced literature of impinging jet atomization with Phase Doppler Anemometry (PDA) are the works of Anderson et al. (1992) and Ryan et al. (1995). PDA is a point measurement technique that can provide very good spatial resolution. Measurements are typically made a few centimeters below the impingement point at the centerline of the spray.

In the works of Anderson et al. (1992) and Ryan et al. (1995) the arithmetic mean drop diameter was observed to decrease with increasing jet velocity. Drop size data was recorded at spatial locations 1.6 and 4.1 cm downstream of the impingement point. $D_{10}$ was used to report the mean drop diameters due to the dynamic range difficulty present. Examining the drop diameter probability density functions in references such as Ryan (1995) shows that the majority of the drops are recorded in the lower end of a dynamic range of approximately 30 – 1300 µm. As a result, the smallest drops were likely not measured and the relatively few large drops recorded distorted the mean drop diameters. The authors recognized this and used the $D_{10}$ mean diameter, rather than $D_{32}$ or *MMD*.

The $D_{10}$ diameters from these studies do not agree well with the established linear stability theories, as outlined by Ryan et al. (1995) and Anderson et al. (2006). However, there exists a possibility that the reason theory and experiments do not agree is because the $D_{10}$ diameter was used instead of $D_{32}$ or *MMD*, as detailed in a technical 28  note by Ibrahim (2009). Ibrahim (2009) argues that the $D_{32}$ diameters obtained by Kang and Poulikakos (1995)
29  using holography agree well with theory.

30  In the study by Anderson et al. (1992) the mean drop axial velocities were also presented at spatial
31  locations 1.6 and 4.1 cm downstream of the impingement point at the centerline of the spray. Drop velocities were
32  measured using Laser Doppler Anemometry (LDA). The drop axial velocity was observed to increase with
33  increasing jet velocity.

34  There is still a paucity of accurate experimental data for comparison with theoretical models. Developments
35  in PDA since the 1990s have made it possible to measure the small drops that may not have been picked up in
36  previous PDA works by Anderson et al. (1992) and Ryan et al. (1995) In the present study, drop size and drop
37  velocity are quantified by the PDA technique but without the dynamic range problem in previous works. In other
38  words, the small drops in the spray were also measured in this experimental work. The high-fidelity experimental
39  data in this work provides clarity for comparisons between experiment and theory. In addition, the measured drop
40  axial velocity distributions presented highlight the polydisperse nature of the spray. Different size drops may be
41  traveling at different velocities a few centimeters after breakup due to interactions with the ambient gas
42  environment.

43

44  **3 Theories of Impinging Jets.**

48  The majority of existing studies in impinging jet atomization have considered the sheet velocity to be equal
49  to the mean jet velocity for theoretical analysis. This is based on the assumption of uniform velocity profile for the
50  jet. Hasson and Peck (1964) used the uniform jet velocity profile assumption to calculate the dimensionless sheet
51  thickness parameter $K^*$, based only on the sheet azimuthal angle $\phi$ and the half-impingement angle $\theta$. Bremond and
52  Villermaux (2006) and Choo and Kang (2007) have suggested that the uniform jet velocity profile is an assumption
53  that may lead to an incorrect derivation of the sheet characteristics.

54  The following part closely follows the work of Bremond and Villermaux (2006). The expression for a
55  liquid jet with a Poiseuille parabolic velocity profile can be transformed based on the elliptical cross section at the
56  point of impingement. A schematic of the sheet (as adapted from Bremond and Villermaux 2006) formed by the
57  impinging jets and the jet cross section is provided in Fig. 2. In this figure: $b^*$ is the dimensionless distance from the
58  center of the jet to the separation point, $q^*$ is the dimensionless radial distance from the separation point, and $\phi$ is the
59  azimuthal angle of the sheet. The terms $b^*$ and $q^*$ are made non-dimensional using the jet radius $R_j$.

61  The conservation equations of mass, momentum, and energy are used to derive the sheet velocity and the
62  sheet thickness parameter using the non-uniform jet velocity profile at high jet Weber numbers (neglecting surface
63  tension) and high jet Froude numbers (neglecting gravity). The jet Froude number is defined as:

$$Fr_j = \frac{U_j}{\sqrt{gd_j}}, \qquad (7)$$

64  where $g$ denotes the acceleration due to gravity. The conservation laws under these assumptions are:

$$\text{Mass: } 2\int_0^{q_j^*} U_j^* \sin\theta \, q^* d\phi \, dq^* = U_s^* K^* d\phi \,, \qquad (8)$$

$$\text{Momentum: } \int_0^{2\pi} \left(U_s^*\right)^2 K^* \cos\phi \, d\phi = -4\pi \cos\theta \int_0^{R_j} U_j^2 r \, dr \,, \qquad (9)$$

$$\text{Energy: } 2\int_0^{q_j} \left(U_j^*\right)^3 \sin\theta \, q^* d\phi \, dq = \left(U_s^*\right)^3 K^* d\phi \,. \qquad (10)$$

65   Here, $U_s^*$ and $K^*$ are the dimensionless sheet velocity and the dimensionless sheet thickness parameters,
66   respectively, under the non-uniform jet velocity profile assumption. The radial location of the jet is symbolized by $r$.
67   The dimensionless location of the jet interface is denoted by $q_j^*$ and is calculated with the expression:

$$q_j^* = \frac{-b^* \cos\phi \, \sin^2\theta + \left(1 - \cos^2\phi \cos^2\theta - (b^*)^2 \sin^2\phi \sin^2\theta\right)^{0.5}}{1 - \cos^2\phi \, \cos^2\theta} \tag{11}$$

68   The integrals in Eqs. 8 and 10 have been analytically solved in Bremond and Villermaux (2006) for the parabolic jet
69   velocity profile. The solutions are noted here as:

$$\int_0^{q_j^*} U_j^* q^* \, dq^* = F\left(b^*, q_j^*, \theta, \phi\right), \tag{12}$$

$$\int_0^{q_j^*} \left(U_j^*\right)^3 q^* \, dq^* = G\left(b^*, q_j^*, \theta, \phi\right). \tag{13}$$

70   Expressions for the dimensionless sheet velocity and the sheet thickness parameter for the non-uniform jet velocity
71   profile are:

$$U_s^* = \frac{U_s}{\overline{U}_j} = \left(\frac{G}{F}\right)^{0.5}, \tag{14}$$

$$K^* = \frac{K}{R_j^2} = \frac{xh}{R_j^2} = \frac{(F)^{1.5}}{(G)^{0.5}}. \tag{15}$$

86   In the above equations: $\overline{U}_j$ is the mean jet velocity, $x$ is the sheet length, and $h$ is the sheet thickness.

87   Instability analysis has been used in literature to predict the breakup of the liquid sheet produced by
88   impinging jets. An exhaustive review on the topic can be found in the work of Sirignano and Mehring (2000).
89   Instability analysis yields an expression known as the dispersion relation, which relates the real part of the growth
90   rate of the unstable wave $\beta_r$ to its wave number $k$. Dombrowski and Johns (1963) assumed the sinuous wave to be
91   dominant and considered a force balance across the liquid sheet in their analysis. This dispersion relation takes into
92   account the viscous, inertial, surface tension, and the aerodynamic pressure forces. The primary cause of the
93   instability was attributed to aerodynamic interaction of the liquid sheet with the surrounding atmosphere. Neglecting
94   the liquid viscosity, the inviscid dispersion relation based on the force balance is:

$$\frac{\beta_r^2 (h/2)^2}{U_s^2} + \frac{2(k)^2 (h/2)^2}{We_s} - 2sk(h/2) = 0. \qquad (16)$$

95    The sheet Weber number $We_s$ is calculated using the expressions:

$$We_s = \frac{\rho_l U_s^2 h/2}{\gamma}. \qquad (17)$$

96    Two different mechanisms exist in literature regarding the formation of drops from the breakup of the
97    liquid sheet. One is a two-stage breakup mechanism proposed by Dombrowski and Johns (1963) and later used by
98    Ryan et al. (1995). In the two-stage breakup mechanism, the sheet first fragments to ligaments and the ligaments
99    then disintegrate to drops. The other is a one-step direct breakup mechanism proposed by Ibrahim and Prekwas
100   (1991), where drops are directly formed from the liquid sheet.

101   Of particular importance for analytical drop size predictions are the sheet thickness parameter $K$ and the
102   sheet velocity $U_s$. In this work, the sheet characteristics derived by Bremond and Villermaux (2006) will be
103   extended to predict drop sizes. Furthermore, their sheet formation analysis will be emulated for the case of the $1/7^{th}$
104   power law turbulent jet velocity profile, and drop sizes will also be predicted based on this jet velocity profile.

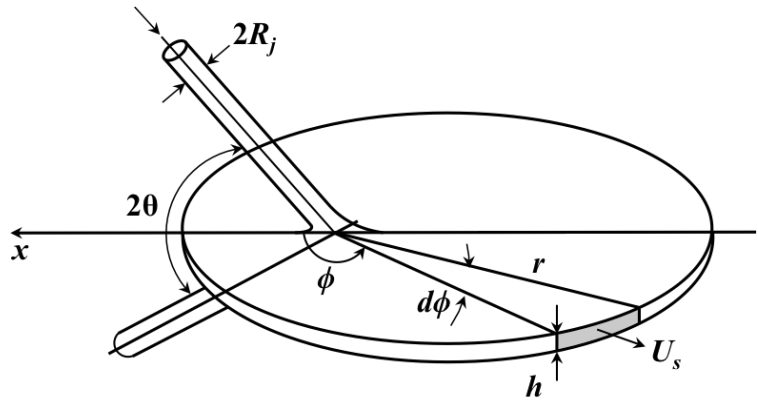

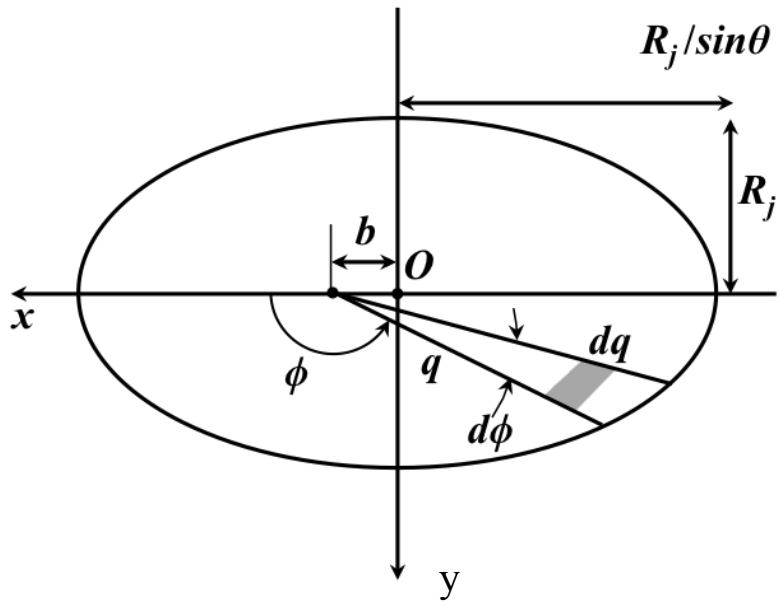

**Figure 2.** Schematic of sheet formation (adapted from Bremond and Villermaux 2006): (a) formation of liquid sheet, (b) elliptical cross-section of inclined jet. Shaded region indicates the differential volume.

## 4 Experimental Facilities.

The unique experimental apparatus used to create atomization in this study was essentially identical to the facility used by Mallory and Sojka (2014) and Rodrigues and Sojka (2014). Figure 3 provides a schematic of the facility. Drop size measurements were made at dimensionless numbers range of: $6{,}820 < We_j < 21{,}900$; $18{,}500 < Re_j < 33{,}100$; and $328 < Fr_j < 588$. Rotational stages were used to specify the impingement angle $2\theta$. Translation stages were used to specify the free jet length-to-orifice diameter ratio $x/d$. Designated tip elements were used to specify the internal length-to-orifice diameter ratio $L/d$ and the orifice diameter $d_0$. The operating pressure was varied to change the mean jet velocity $\overline{U_j}$. The geometric parameters were: $d_0 = 0.686$ mm, $2\theta = 100°$, $x/d = 60$, $L/d = 20$. Deionized water was used as the test liquid and the literature values are used for liquid density ($\rho_l = 1000$ kg/m$^3$) and liquid viscosity ($\mu_l = 0.001$ Pa-s). The ambient gas was air at atmospheric temperature and pressure; literature values were used for gas density ($\rho_g = 1.2$ kg/m$^3$) and gas viscosity ($\mu_g = 1.85$E-05 Pa-s). The liquid/gas surface tension was taken to be the literature value of water/air at atmospheric conditions ($\gamma = 0.0728$ N/m). The mean jet velocity was measured using a stopwatch for flow rate test durations of 30 seconds. Flow rate measurements were repeated three times to ensure statistically significant mean jet velocity measurements. The flow exiting the orifice is believed to be turbulent due to the high Reynolds numbers. A discharge coefficient of $0.70 \pm 0.02$ was measured for the orifices. Cavitation was not visibly observed for the test conditions used in this work, despite the high jet velocities. This can be attributed to the fairly high $L/d$ ratio and the rather small orifice diameter.

A PDA system was used to obtain measurements of drop size and drop velocity. Phase Doppler Anemometry (PDA) measures the drop diameter by measuring the phase difference between Doppler signals from two different detectors. The drop velocity can also be measured by the PDA system by measuring the frequency of the Doppler bursts, which are obtained by converting the optical signals from the detectors. These Doppler bursts have a frequency that is linearly proportional to the drop velocity. More details about the theory behind PDA may be found in Albrecht et al. (2003). The PDA receiver with a 310 mm focal length lens was oriented 30° from the laser beams produced by the PDA transmitter. The PDA transmitter emits a pair of Helium-Neon and Nd:YAG laser beams and was used with a 400 mm focal length lens. Figure 4 provides a schematic of the set-up for the PDA optical diagnostics.

The PDA was configured with particular hardware and software settings to yield high validation and data rates. All measurements were taken at the centerline of the spray at 5 cm below the point of impingement. The injector pair was mounted on traverse stages that enable movement in all three spatial dimensions for alignment. 50,000 data points were collected for each test. A curtain of air was placed in front of the PDA receiver using an air-knife to prevent drops from some test conditions from landing on the lens. Measurements were taken with and without the air curtain at a baseline condition for comparison to ensure that the air does not have an effect on the measurements.

The PDA optical configuration enables a drop size measurement range between approximately 2.3 to 116.2 μm by using the selected aperture plate (Mask B). This was chosen as the optimal measurement range over two other options of 1.4 to 71.6 μm (Mask A) and 5.5 to 276.8 μm (Mask C). A truncation analysis showed that Mask C did not detect the small drops that were detected by the other two Masks. Mask B was not only observed to detect the small drops that were detected by Mask A, but also detect the larger drops that were not detected by Mask A. Further details on the aperture plate selection can be found in Rodrigues (2014). In direct contrast, recall that the drop size measurement range used by Anderson et al. (1992) and Ryan et al. (1995) was about 30-1300 um. The smallest drops were certainly not detected in the previous works.

The experimental setup for imaging the impinging jets is shown in Fig. 5. The back illumination was provided by a double-pulsed Nd:YAG laser beam. The pulse width was 6 ns, which is short enough to freeze the motion of the liquid. The laser beam was first expanded and then projected to a diffuser, thus significantly reducing the coherence of the laser beam. Accordingly, no interference patterns were observed in the recorded images. In addition, the effect of the speckle noise on image quality was unnoticeable due to the relatively small speckle size. A CCD camera was focused at the plane of the two impinging jets, which was back-illuminated by the light produced by the diffuser.

Table 1 provides the percent uncertainty for the experimental work. The percent uncertainty for the operating conditions was calculated using the method delineated in Kline and McClintock (1953). Mean drop diameters ($D_{10}$, $D_{32}$, $MMD$) and mean axial drop velocity ($U_{z\text{-}mean}$) uncertainties were investigated by taking 5 repeated measurements at a baseline operating condition. The coefficient of variation, the ratio of one standard deviation to the mean, was used as the uncertainty for the PDA measurements. Further details on calculating the uncertainty can be found in Rodrigues (2014).

166

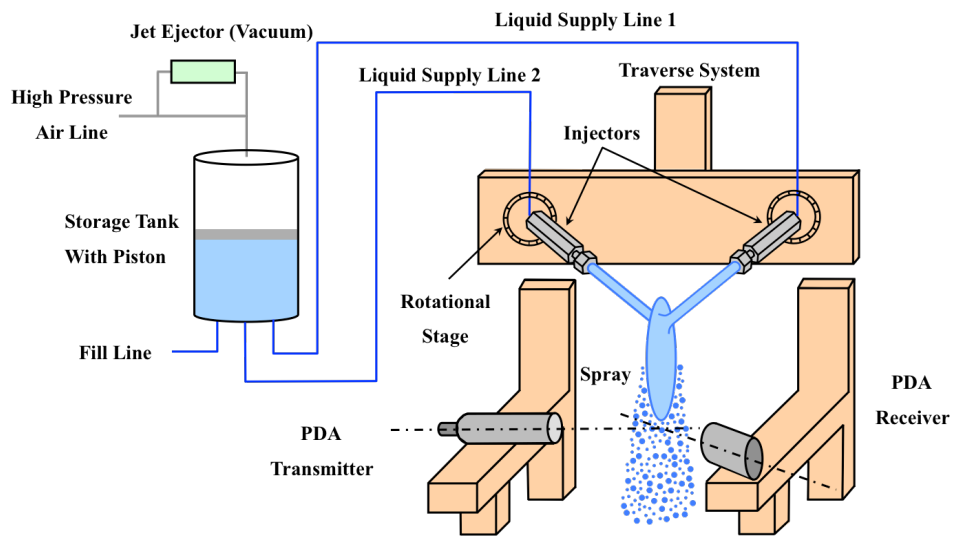

167

**Figure 3.** Experimental apparatus schematic.

169

170

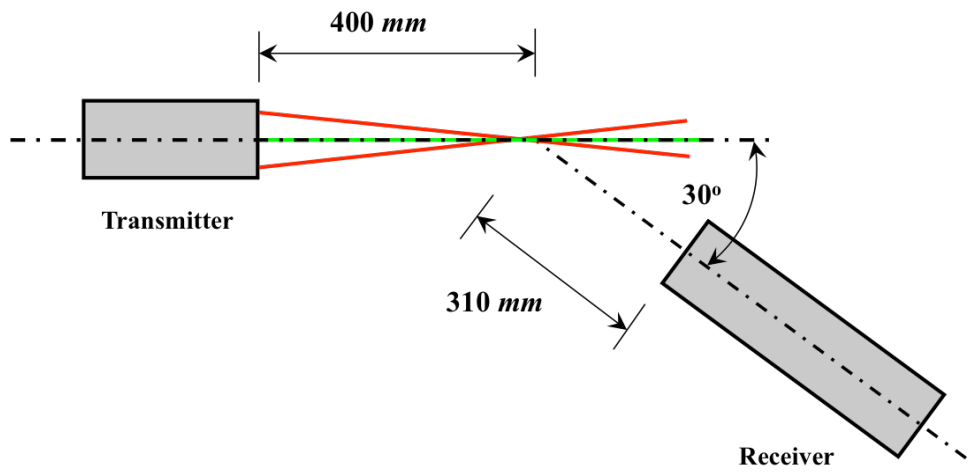

171 **Figure 4.** PDA set-up schematic.

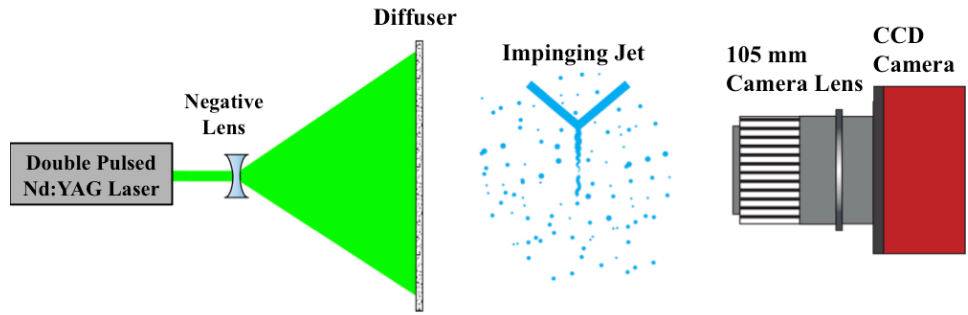

172

173 **Figure 5.** Schematic for imaging system.

174

175          **Table 1.** Experimental uncertainty for presented experimental data.

| Quantity | Uncertainty |
|---|---|
| $We_j$ | 4.5% |
| $D_{10}$ | 5.9% |
| $D_{32}$ | 1.3% |
| MMD | 0.9% |
| $U_{z\text{-}mean}$ | 2.9% |

176

177

178

179

180

## 5. Results and Discussion.

Figure 6 presents a selection of number, surface area, and volume probability density functions (pdf) along with the corresponding spray patterns for increasing jet Weber numbers. It was observed that the sheet breaks up a short distance from the point of impingement. An increase in the denseness of the spray with increasing jet Weber numbers was also observed, which was consistent with the greater data rate in the PDA measurements at higher $We_j$. The number pdf was not observed to vary significantly with increasing jet Weber numbers. This is likely due to the high quality of atomization in the impact wave regime. However, variations were observed in the surface area and volume pdfs, which indicates that a smaller portion of larger drops is present at higher jet Weber numbers.

Experimental values of $D_{10}$, $D_{32}$, and $MMD$ are presented as a function of the jet Weber number in Fig. 7. The $D_{10}$ drop diameter was not significantly affected by an increase in $We_j$, due to the large number of small drops spanning the range of jet Weber number. However, the decrease in the $D_{32}$ mean diameter indicates that the fineness of the spray was improved by the enhancement of the inertial force. $MMD$ was also observed to decrease with increasing $We_j$. In this work, $MMD$ is used for comparison with analytical diameter predictions because the instability model is based on a momentum balance and therefore must have a mass/volume foundation to it.

In Bremond and Villermaux (2006), the momentum equation (Eq. 9) was numerically solved in order to determine the value for $b^*$, the dimensionless distance from the center of the jet to the separation point. It was determined that $b^* = 0.68/\tan\theta$ for the parabolic jet velocity profile. For a $1/7^{th}$ power law turbulent jet velocity profile, the dimensionless velocity profile is given by:

$$U_{j,t}^* = \frac{U_{j,t}}{\overline{U}_j} = 1.22\left(1 - \frac{r}{R_j}\right)^{\frac{1}{7}} \tag{18}$$

In the above expression $U_{j,t}$ is the dimensional turbulent jet velocity and $r$ is the radial location of the jet. The maximum velocity of 1.22 times the mean jet velocity occurs at $r = 0$ (jet centerline) and the minimum velocity of 0 occurs at $r = R_j$ (jet edge). Equation 9 was numerically solved for the turbulent jet velocity profile for a half-impingement angle of 50° to yield $b^* = 0.755$. Using this value for the dimensionless distance from the center of the jet to the separation point, Eqs. 11 – 15 were solved for the turbulent jet velocity profile. Table 2 presents the values for $b^*$, ratio of sheet velocity to mean jet velocity $\alpha$, and dimensionless sheet thickness parameter $K^*$ for the uniform, 205  parabolic, and turbulent jet velocity profiles at the corresponding experimental conditions (impingement angle $2\theta$ =
206  100° and sheet azimuthal angle $\phi = 0°$).

207  Squire (1953) used the long wave approximation ($kh/2 \ll 1$) to solve the dispersion relation for the
208  maximum growth rate $\beta_{r,max}$ and its corresponding wave number $k_{max}$. These are expressed as:

$$\beta_{r,\max} = \frac{\rho_g U_s^2}{\sqrt{2\rho_l \gamma h}}, \tag{19}$$

$$k_{\max} = \frac{\rho_g U_s^2}{2\gamma}. \tag{20}$$

209  In the one step breakup mechanism by Ibrahim and Prekwas (1991) it is assumed that sheet breakup is due
210  to the growth of waves at the maximum amplification rate. The resulting drop size is determined as half of the
211  wavelength of the fastest growing waves $\lambda_{max}$. Since $k_{max} = 2\pi/\lambda_{max}$:

$$d_D = \frac{\pi}{k_{\max}}. \tag{21}$$

212  Equations 20 and 21 can be combined and the dimensionless drop diameter can be expressed in terms of the jet
213  Weber number $We_j$, ratio of sheet velocity to mean jet velocity $\alpha$, and ratio of ambient gas density to liquid density $s$.
214  The resulting analytical expression for the drop diameter is dependent on the assumed jet velocity profile:

$$\frac{d_D}{d_0} = \frac{2\pi}{\alpha^2 s We_j}. \tag{22}$$

215  For the case of the uniform jet velocity profile, $\alpha = 1$, Eq. 22 is reduced to the expression proposed by
216  Ibrahim and Przekwas (1991):

$$\frac{d_D}{d_0} = \frac{2\pi}{s We_j}. \tag{23}$$

217  One obvious disadvantage of the one-step breakup mechanism is the lack of dependency on the sheet azimuthal
218  angle and half-impingement angle.

219    A condition for breakup needs to be first established for the two-step breakup mechanism. Dombrowski
220 and Johns (1963) based the criterion for sheet breakup on the following empirical relation:

$$\int_0^{x_b} \frac{\beta_{r,\max}}{U_s} dx = 12, \quad (24)$$

221 where $x_b$ is the sheet breakup length. The choice of the constant 12 is based on experimental observations by Weber
222 (1931). This expression is for an attenuating sheet and therefore the growth rate must be integrated to predict the
223 breakup length. The sheet breakup length is used with the sheet thickness parameter $K$ to calculate the sheet
224 thickness at breakup $h_b$:

$$K = x_b h_b. \quad (25)$$

225 Dombrowski and Johns (1963) then calculated the ligament diameter $d_L$ using the expression:

$$d_L = \sqrt{\frac{4 h_b}{k_{\max}}}. \quad (26)$$

226    In practice curved ligaments break-off from the sheet, as observed in the images of Ryan et al. (1995)
227 amongst others. However, the curvature of the ligaments is large compared to their diameter. As argued by
228 Dombrowski and Johns (1963), this permits the use of the Plateau-Rayleigh analysis to calculate the drop diameter
229 $d_D$:

$$d_D = 1.88 d_L. \quad (27)$$

230    Combining the expressions in Eqs. 19, 20, 24-27 the dimensionless drop diameter can be expressed in
231 terms of the jet Weber number $We_j$, ratio of sheet velocity to mean jet velocity $\alpha$, dimensionless sheet thickness
232 parameter $K^*$, and ratio of ambient gas density to liquid density $s$:

$$\frac{d_D}{d_0} = \frac{1.14 (K^*)^{1/3}}{\alpha^{2/3} s^{1/6} We_j^{1/3}}. \quad (28)$$

233    For the case of the uniform jet velocity profile, $\alpha = 1$, Eq. 28 is reduced to the expression proposed by Ryan
234 et al. (1995):

$$\frac{d_D}{d_0} = \left[\frac{2.62}{12^{1/3}}\right] s^{-1/6} [We_j f(\theta)]^{-1/3}, f(\theta) = \frac{(1-\cos\phi\cos\theta)^2}{\sin^3\theta} \ . \tag{29}$$

235  Note that for the case of the uniform jet velocity profile, $K^* = 1/f(\theta)$.

236  In this experimental work the jets did not have a parabolic profile inside the orifice and instead likely had a
237  turbulent profile due to the large jet Reynolds numbers. However, as argued by Bremond and Villermaux (2006),
238  considering the parabolic profile is a worthwhile investigation of another limit to complement the investigated
239  models of prior researchers who studied the uniform velocity profile limit. Diameter predictions from parabolic,
240  turbulent, and uniform velocity profiles are presented in this work.

243  Figure 8a presents a comparison of the experimental *MMD* and predicted diameters using the one-step
244  breakup process for the parabolic, turbulent, and uniform velocity profiles. It can be clearly observed that
245  predictions from the uniform profile used by Ibrahim and Przekwas (1991) drastically over-predict the experimental
246  *MMD*. This can be attributed to the assumption of uniform velocity profile, which leads to the sheet velocity being
247  equal to the jet velocity. Diameter predictions closer to the experimental data were observed for the turbulent and
248  parabolic jet velocity profiles. However, note that the trends from all three models that use this one-step breakup
249  process did not agree with the trends of the experimental *MMD*. This is because Eq. 22 does not satisfactorily
250  capture the non-linearity in the atomization process.

251  A comparison of the experimental *MMD* to the predicted diameters given by the two-step breakup
252  mechanism with the attenuating sheet for parabolic, turbulent, and uniform jet velocity profiles is presented in Fig.
253  8b. Once again it can be clearly observed that predictions from the uniform profile used by Ryan et al. (1995)
254  significantly over-predict the experimental *MMD*. Diameter predictions from the turbulent jet velocity profile offer a
255  closer comparison to the experiential data, and predictions from the parabolic jet velocity provide almost an exact
256  match.

257  An empirical correlation with an $R^2 = 0.97$ for drop diameter based on the experimental *MMD* is:

$$\frac{d_D}{d_0} = \frac{0.325}{s^{1/6} We_j^{0.201}} \ . \tag{30}$$

258   It should be noted that the only density ratio tested for this experimental work was air to water, both at atmospheric
259   conditions. Therefore, the Ryan et al. (1995) exponent of -1/6 for the density ratio is assumed here. The jet Weber
260   number range tested was from 6,820 to 21,900. The exponent for $We_j$ presented here (0.201) is lower than those of
261   the one-step breakup mechanism (1) and two-step breakup mechanism (1/3). This indicates a weaker dependency on
262   $We_j$ according to the experimental data. Therefore, although trends from the two-step breakup mechanism agree
263   better than trends from the one-step breakup mechanism, further work is needed to incorporate more of the non-
264   linearity.

265   Higher mean drop velocities were observed with an increase in the jet Weber number, due to the increase in
266   the inertial force. Experimental values of the drop axial velocity are compared with a model based on drop ballistics
267   theory. Equation 31 provides the standard governing ordinary differential equation used to calculate the drop
268   velocity $U_d$ based on the axial position $x$. The equation balances the inertial force with the gravitational force and
269   drag force:

$$\frac{dU_d}{dx} = \frac{g}{U_d} - \frac{3C_{drag}}{4d_D} s U_d. \tag{31}$$

270   In the above expression, $C_{drag}$ is the drag coefficient and is calculated using the criteria:

$$C_{drag} = \begin{cases} 24/Re_D, & Re_D < 2 \\ 18.5/Re_D^{0.6}, & 2 < Re_D < 1000 \\ 0.44, & 1000 < Re_D \end{cases} \tag{32}$$

271   The drop Reynolds number is calculated using the expression:

$$Re_D = \frac{\rho_g U_D d_D}{\mu_g}. \tag{33}$$

272   The drop diameter used to calculate the Reynolds number of the drop $Re_D$ was the experimentally determined $D_{32}$.
273   $D_{32}$ was used as the mean diameter because the surface area and volume of the drop are the basis for the drop
274   ballistics model. Literature values for air at atmospheric conditions were used for the gas density $\rho_g$ and gas
275   viscosity $\mu_g$.

277  The initial velocity of the drop $U_d$ was assumed to be the velocity of the sheet at $\phi = 0°$. The sheet velocity
278  is dependent on the jet velocity profile assumed (parabolic, turbulent, or uniform). The drop Weber number $We_D$ is
279  calculated using the expression:

$$We_D = \frac{\rho_g U_D d_D}{\gamma}. \tag{34}$$

280  The experimentally determined $D_{32}$ was also used as the drop diameter to calculate $We_D$. The calculated drop Weber
281  numbers ($0.19 < We_D < 0.51$) were well below the established regimes for secondary atomization (Guildenbecher et
282  al. 2009). Therefore, surface tensions effects connected with the drop deformation were neglected in this drop
283  ballistics model.

284  The initial location of the drop was assumed to be the sheet breakup length. This was calculated using a
285  phenomenological relation (Anderson et al. 2006):

$$\frac{x_b}{d_0} = 13.56 \ (We_j s^2)^{-0.102}. \tag{35}$$

286  Experimental diameter data and a phenomenological relation for breakup length were used instead of predictions
287  based on instability analysis in order to ensure that deficiencies in the instability theory were not introduced into the
288  drop ballistics model.

289  Comparisons between theoretical predictions based on the three jet velocity profiles and experimental data
290  are presented in Fig. 9. It should be noted that the experimental uncertainty is within the symbol, and the vertical
291  bars shown are the root mean square (RMS) of velocity. The predictions with the turbulent jet velocity profile agree
292  best with the experimental data. Predictions with the parabolic jet velocity profile over-predict the drop velocities
293  and predictions with the uniform velocity profile slightly under-predict the drop velocities. These observations are
294  directly related to the initial velocity of the drop, which in turn ultimately depends on the ratio of sheet velocity to
295  mean jet velocity $\alpha$. Velocity predictions based on the turbulent $\alpha = 1.08$ agree best with the experimental drop
296  velocities, since the turbulent jet velocity profile most closely corresponds with the experiment condition.

297  Due to the large root mean square values, a more detailed look at the drop velocities is required. The pdfs
298  of drop velocities at different jet Weber numbers are shown in Fig. 10, where relative velocity obtained by 299  subtraction of the mean drop velocity is presented. At higher $We_j$ it was observed that there is a shorter/flatter peak
300  for the drop velocity with the greatest probability density. The distribution was also observed to become increasingly
301  wider. This indicated that at the higher jet Weber numbers, the drop velocities in the spray become increasingly
302  polydispersed.

303

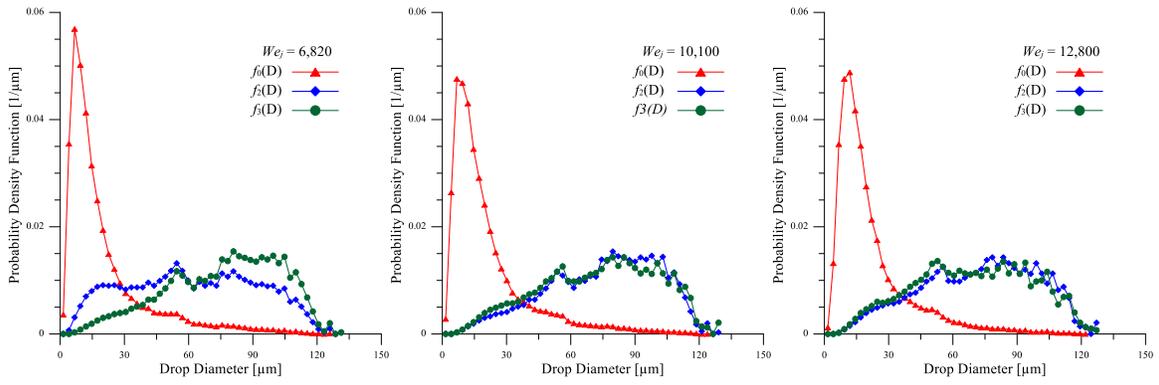

(a)             (b)             (c)

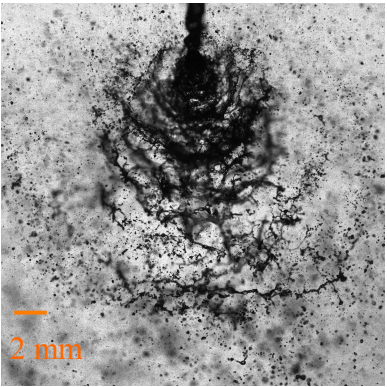
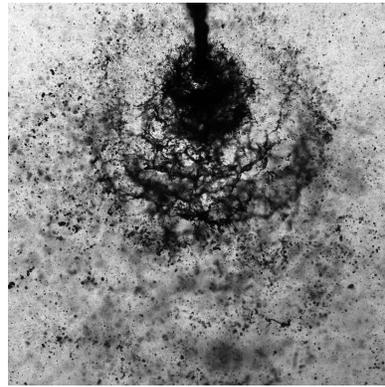
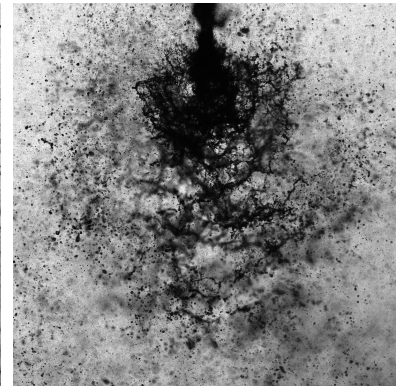
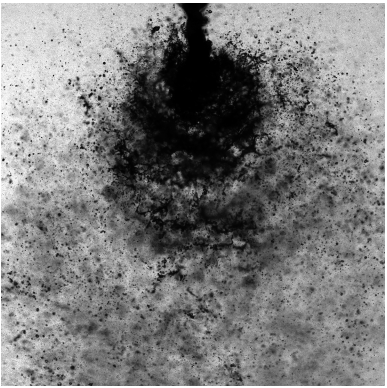
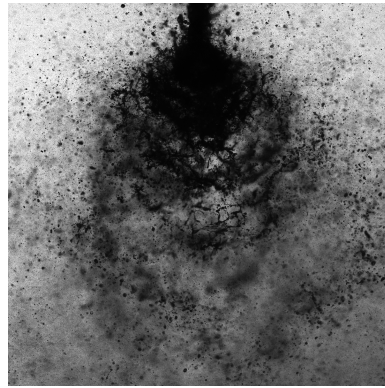
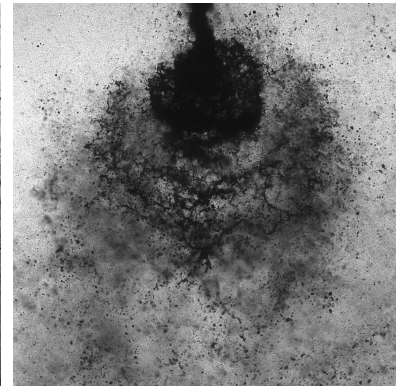

(d)             (e)             (f)

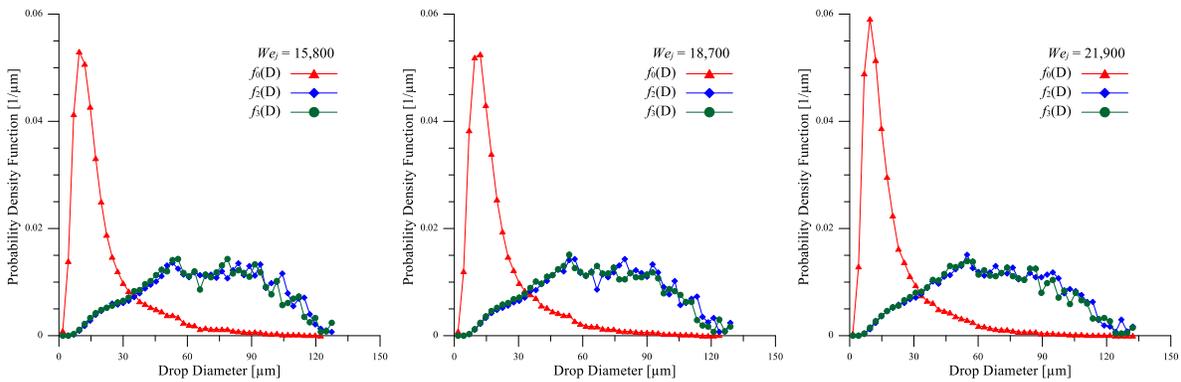

**Figure 6.** Number, area, and volume pdfs versus drop diameters with corresponding spray patterns for: (a) $We_j = 6,280$, (b) $We_j = 10,100$, (c) $We_j = 12,800$, (d) $We_j = 15,800$, (e) $We_j = 18,700$, (f) $We_j = 21,900$.

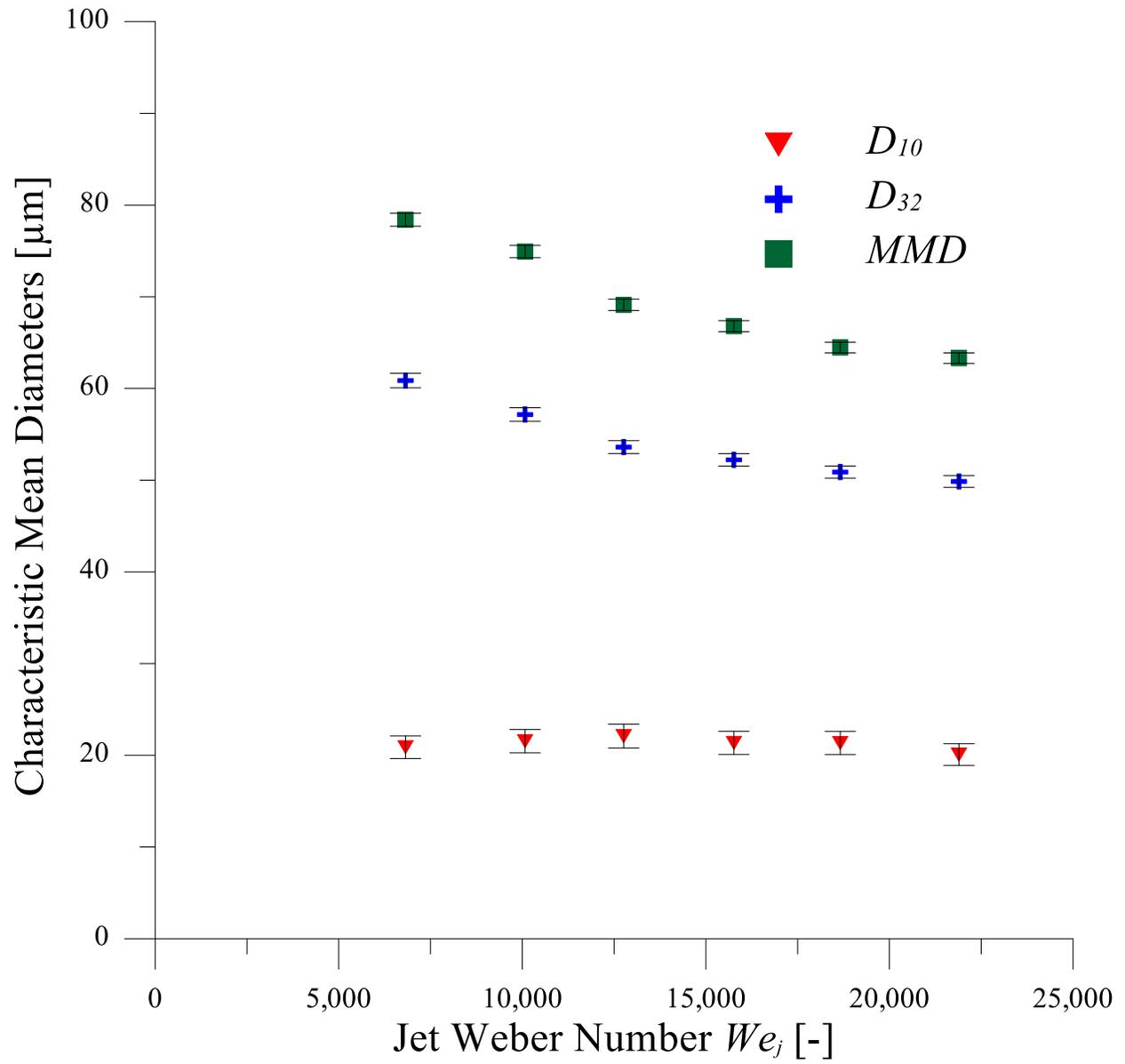

**Figure 7.** Experimentally measured $D_{10}$, $D_{32}$, and *MMD* versus jet Weber number.

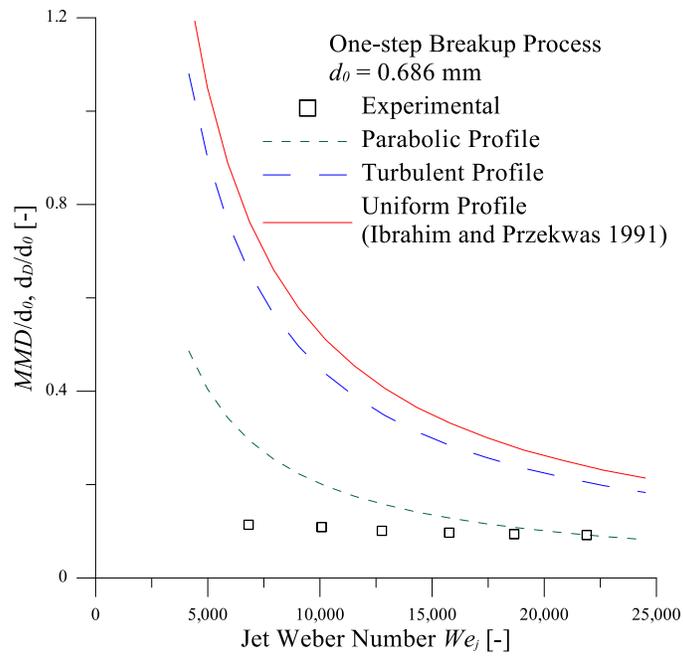

(a)

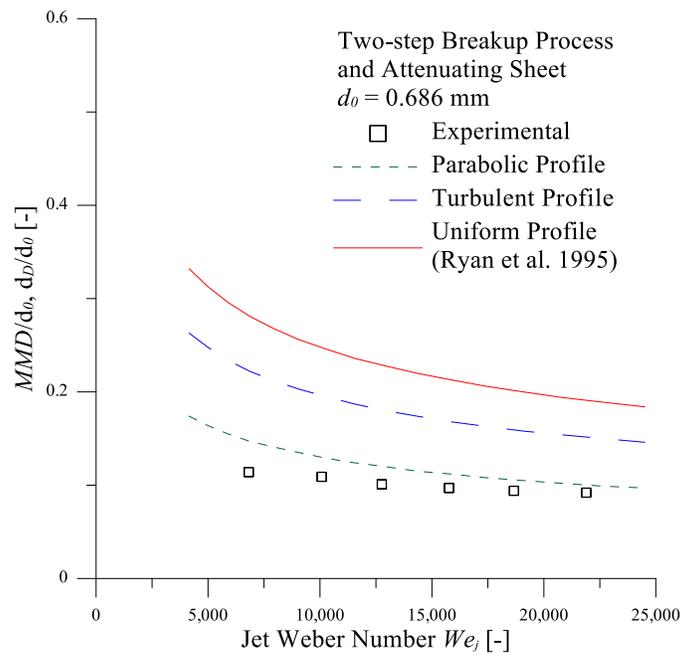

(b)

**Figure 8.** Dimensionless mass median diameter and predicted drop diameters versus jet Weber number for: (a) one-step breakup mechanism, (b) two-step breakup mechanism with attenuating sheet.

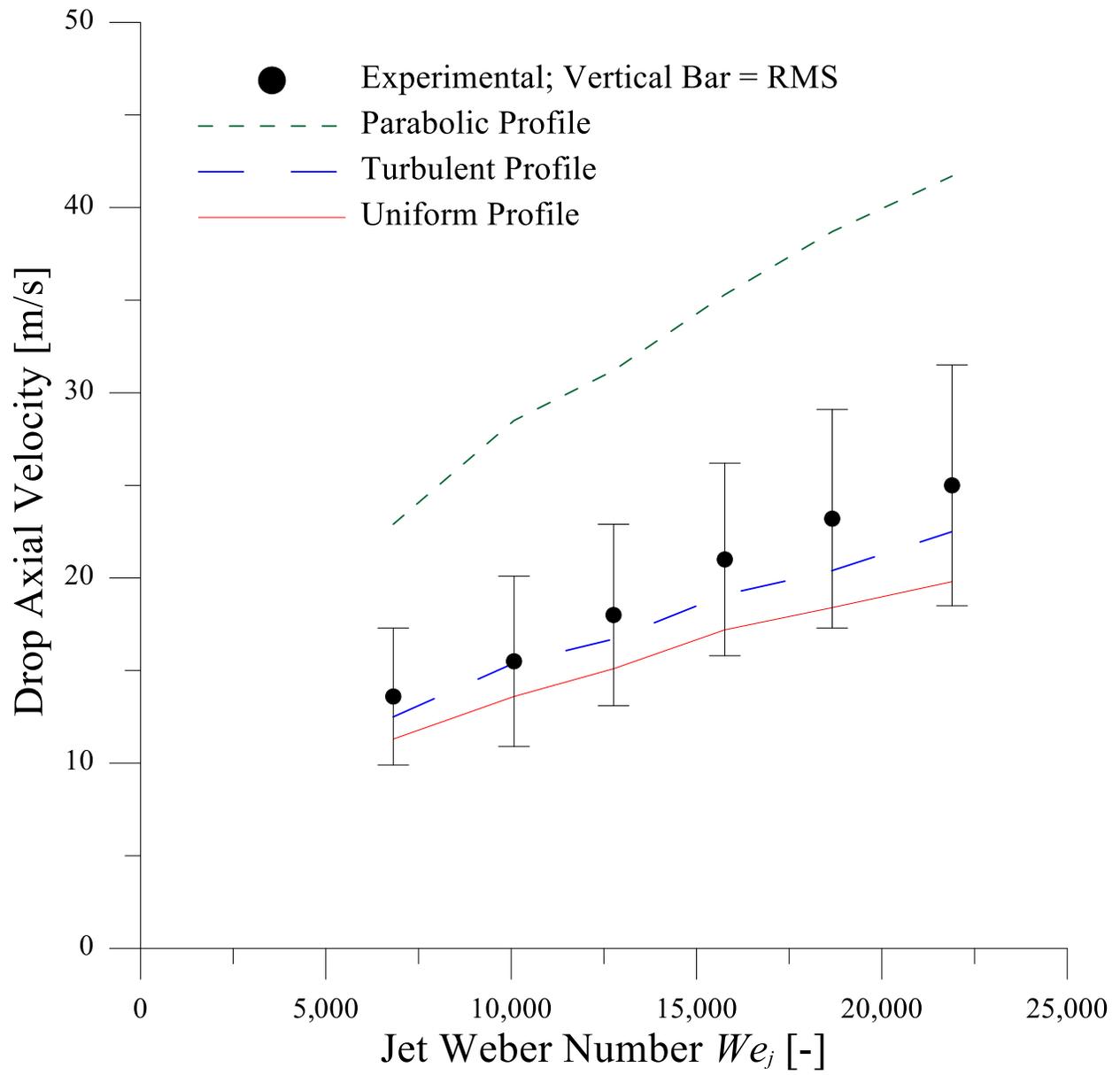

**Figure 9.** Measured and predicted drop axial velocity versus jet Weber number.

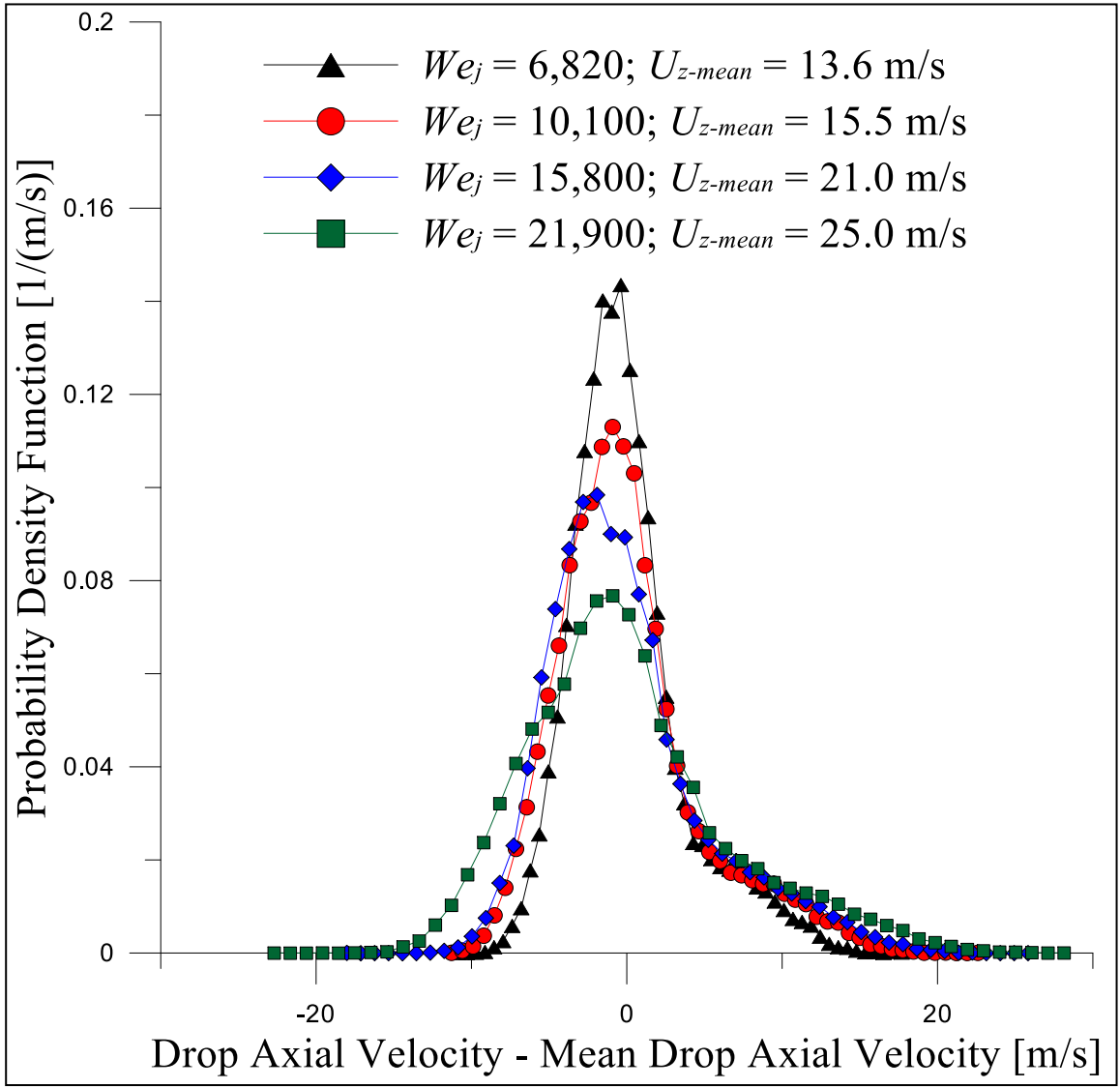

**Figure 10.** Measured axial velocity pdf versus drop axial velocity minus mean drop axial velocity.

**Table 2.** Calculated dimensionless parameters for three jet velocity profiles at $2\theta = 100°$, $\phi = 0°$.

| Jet Velocity Profile | $b^*$ | $\alpha$ | $K^*$ |
|---|---|---|---|
| Parabolic | 0.571 | 1.61 | 1.34 |
| Turbulent | 0.755 | 1.08 | 2.07 |
| Uniform | 0.839 | 1.00 | 3.52 |

**6 Conclusions.**

In the experimental portion of this work, measurements were presented for the means and distribution of drop diameter and drop velocity in the impact wave regime. A modern PDA system that has the capacity to measure the smallest drops in the spray was used in order to correct the dynamic range problem from the previous studies. Number, surface area, and volume pdfs provided an indication of the variety of drop sizes in the spray. The $D_{32}$ and *MMD* drop diameters were observed to decrease with increasing $We_j$, which was due to a decrease in the number of larger drops at higher jet Weber numbers. The mean drop velocity was observed to increase with increasing $We_j$. This was due to the increased inertial force at higher jet Weber numbers. Probability density functions of drop velocity were presented and a wider distribution for drop velocities was observed at higher $We_j$. This indicates an enhancement of the polydisperse nature of the spray as the inertial force was increased.

The importance of the assumed jet velocity profile on analytical drop diameter predictions was illustrated in this work. Analytical drop diameter models in literature, which assume a uniform jet velocity profile, showed poor agreement with experimental *MMD* for both one-step and two-step breakup mechanism. Analytical expressions that depend on parameters based on the assumed jet velocity profile were presented for both the one-step and two-step breakup mechanism. Predictions based on the parabolic and turbulent jet velocity profiles using the two-step breakup mechanism showed close agreement to the experimental data. For all jet velocity profiles, predictions using the one-step breakup mechanism showed poor agreement to the experimental data. Velocity predictions from the drop ballistics model using the turbulent jet velocity profile agreed best with the experimentally measured mean drop velocities.

**Appendix A**

For the assumption of a parallel-sided sheet, instead of an attenuating sheet, the sheet thickness is constant. Therefore, as outlined in Senecal et al. (1999), the following expression can be used as a condition for breakup:

$$\frac{x_b \beta_{r,\max}}{U_s} = 12 . \tag{A1}$$

In keeping with the long-wave approximation ($kh/2 \ll 1$), an analytical expression can also be derived for drop diameter. This expression depends on the jet Weber number $We_j$, ratio of sheet velocity to mean jet velocity $\alpha$, dimensionless sheet thickness parameter $K^*$, and ratio of ambient gas density to liquid density $s$:

$$\frac{d_D}{d_0} = \frac{1.30(K^*)^{1/3}}{\alpha^{2/3} s^{1/6} We_j^{1/3}} . \tag{A2}$$

Figure A1 presents a comparison of the experimental *MMD* to the predicted diameters given by the two-step breakup mechanism with the parallel-sided sheet for parabolic, turbulent, and uniform jet velocity profiles. The parallel-sided sheet assumption is useful for predictions for viscous Newtonian and non-Newtonian liquids, where the invisid assumption cannot be justified.

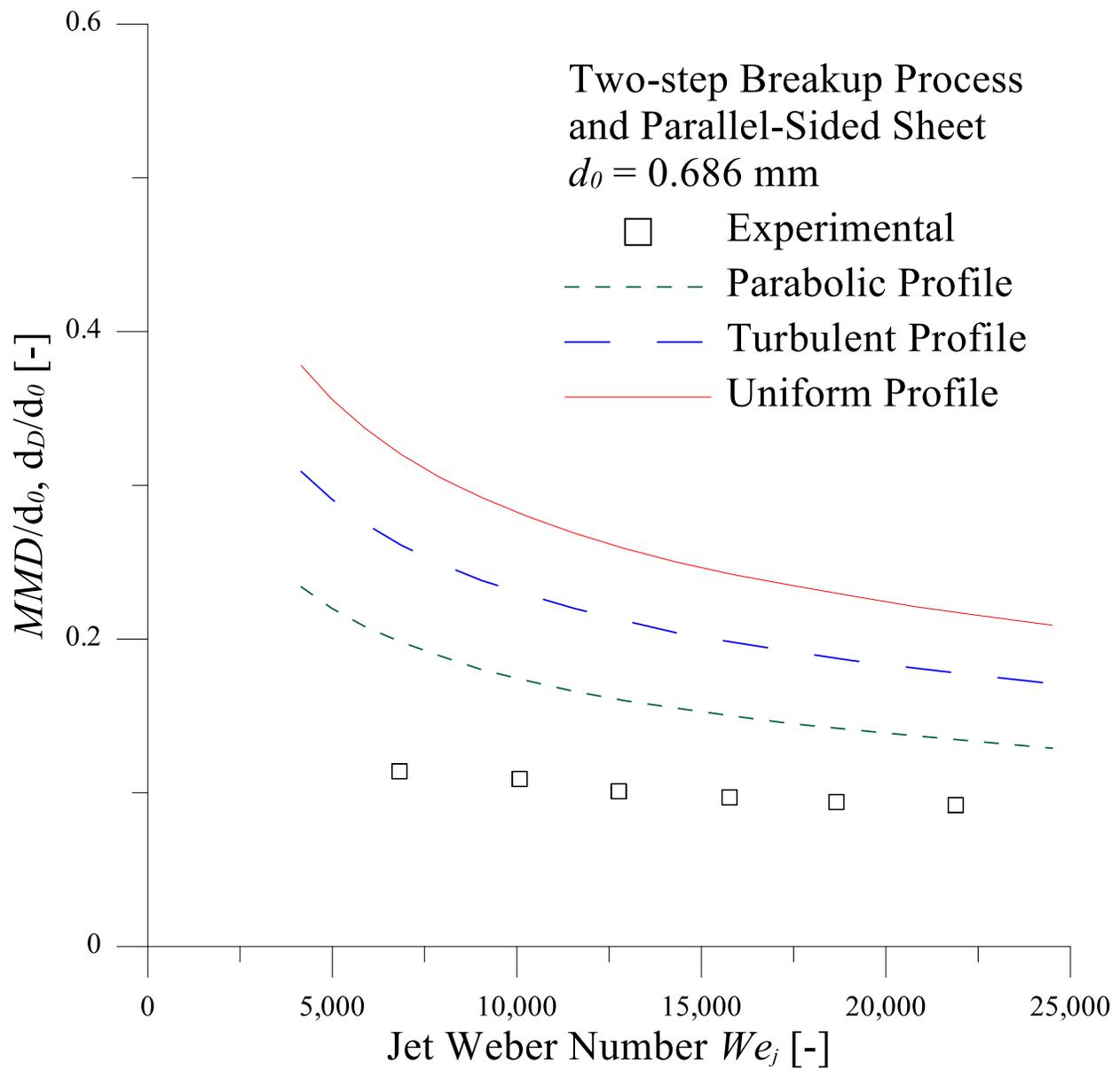

Figure A1 Dimensionless mass median diameter and predicted drop diameters versus jet Weber number for two-step breakup mechanism with parallel-sided sheet.

**Acknowledgements.**

The research presented in this paper was made possible with the financial support of the U.S. Army Research Office under the Multi-University Research Initiative grant number W911NF-08-1-0171. N.S. Rodrigues thanks Prof. Jennifer Mallory for helpful feedback, Dr. Ariel Muliadi for assistance with PDA configuration, and Prof. William Anderson for fruitful discussions.